\documentclass[rnote,traditabstract]{aa}
\usepackage[authoryear]{natbib} \bibpunct[]{(}{)}{;}{a}{,}{,}
\usepackage[latin2]{inputenc} \usepackage[IL2]{fontenc}
\usepackage[czech,english]{babel}
\usepackage{amsmath} 
\usepackage{amssymb}
\usepackage{graphicx} \graphicspath{{\jobname/}}

\newcommand{\sssm}[1]{\scriptscriptstyle\mathrm{#1}}

\newcommand{\msun}{\ensuremath{\mathrm{M}_{\odot}}}
\newcommand{\mdot}{\ensuremath{\dot{M}}}

\makeatletter
\@twosidefalse
\@mparswitchfalse
\makeatother

\begin{document}
\bibliographystyle{aa}

\title{Gravitational excitation of high frequency QPOs}
\titlerunning{Gravitational excitation of QPOs}

\author{Zden\v{e}k Stuchl\'{\i}k\inst{1}
      \and
      Sushan Konar\inst{2}
      \and
      John C. Miller\inst{3,4}
      \and
      Stanislav Hled\'{\i}k\inst{1}}
\authorrunning{Z. Stuchl\'{\i}k et al.}

\offprints{Stanislav Hled\'{\i}k}

\institute{Institute of Physics,
      Faculty of Philosophy and Science, Silesian University in Opava\\
      Bezru\v{c}ovo n\'am. 13, CZ-74601 Opava, Czech Republic\\
      \email{zdenek.stuchlik@fpf.slu.cz}, \email{stanislav.hledik@fpf.slu.cz}
      \and
      Centre for Theoretical Studies, Indian Institute of Technology,
      Kharagpur 721\,302, India\\
      \email{sushan.konar@googlemail.com}
      \and
      SISSA, International School for Advanced Studies and INFN\\
      Via Beirut 2--4, I-34014, Trieste, Italy\\
      \email{miller@sissa.it}
      \and
      Department of Physics (Astrophysics), University of Oxford,
      Keble Road, Oxford OX1 3RH, England\\
      \email{jcm@astro.ox.ac.uk}}

\date{Submitted August 4, 2008}

\abstract
{ We discuss the possibility that high-frequency QPOs in neutron-star binary 
  systems may result from forced resonant oscillations of matter in the 
  innermost parts of the accretion disc, excited by gravitational perturbations 
  coming from asymmetries of the neutron star or from the companion star. We 
  find that neutron-star asymmetries could, in principle, be effective for 
  inducing both radial and vertical oscillations of relevant amplitude while 
  the binary companion might possibly produce significant radial oscillations 
  but not vertical ones. Misaligned neutron-star quadrupole moments of a size 
  advocated elsewhere for explaining limiting neutron star periods could be 
  large enough also for the present purpose.}

\keywords{neutron stars -- binary systems -- QPOs -- accretion discs}

\maketitle

\section{Introduction}\label{intro}

Quasi-periodic oscillations of \mbox{X-ray} brightness (QPOs) have 
been observed in a number of accreting binary systems containing 
compact objects, both with neutron stars \citep[see][ for a 
review]{Kli:2000:ARASTRA:,Bar-Oli-Mil:2005:MONNR:} and with black 
holes \citep{Rem-McCli:2006:ARASTRA:}. They can have low frequencies 
($\mathrm{Hz}$) or high frequencies ($\mathrm{kHz}$). The observed 
$\mathrm{kHz}$ frequencies are comparable with the Keplerian and 
epicyclic frequencies in the inner parts of the accretion disc 
\citep{Tor:2005:ASTRN:} and the question arises of whether they might 
be associated with forced resonant oscillations of the inner disc 
material. In order to initiate these, some perturbation mechanism 
would be required. Here, we focus on neutron-star systems and 
investigate the possibility that gravitational perturbations caused 
either by the binary companion or by asymmetries of the neutron star 
might provide this mechanism. (The case of the binary companion could 
be relevant also for black hole systems.) We note that perturbations 
coming from the neutron star can only be relevant for the innermost 
parts of the disc (because of the rapid fall-off of the force with 
distance) and hence are mainly associated with the picture for 
$\mathrm{kHz}$ QPO frequencies rather than with that for the lower 
frequencies. It is necessary that the mechanism should be a resonant 
one since otherwise the response produced would certainly be much too 
small to be of interest.

The two types of perturbation (from the neutron star and from the 
binary companion) clearly induce different behaviours: the frequency 
of the varying force arising from the influence on the disc of the 
binary companion is essentially equal to the disc rotation-frequency, 
whereas the main frequency of the force caused by asymmetries of the 
neutron star is equal to the difference between the rotation 
frequencies of the disc and the neutron star \citep{Petri:2006:ApSS:}. 
Resonance occurs at those points where one of the intrinsic 
oscillation frequencies of the disc matter coincides with the forcing 
frequency. Following the discussion by \citet{Lan-Lif:1976:Mech:} for 
test particle motion (which would need to be modified for fluid 
elements), the growth in amplitude of the oscillations of the particle 
within the linear regime of forced resonance is given by
 \begin{equation}
   a(t) = \frac{f_{\mathrm{p}}}{2m_0\omega}\,t\ , \label{resonance}
\end{equation}
 where $f_{\mathrm{p}}$ is the amplitude of the variations of the force, 
$\omega$ is the frequency and $m_0$ is the mass of the particle. This linear 
regime ends when the oscillation amplitude $a(t)$ becomes large enough so that 
non-linear phenomena and/or dissipative processes become relevant. Note that 
$a(t)$ grows linearly with time in this regime and so can become quite large 
even when the variations in the perturbing force are small.

\section{Gravitational perturbing force}\label{graforce}

\begin{figure}[t]
\begin{center}
\includegraphics[width=.95\linewidth]{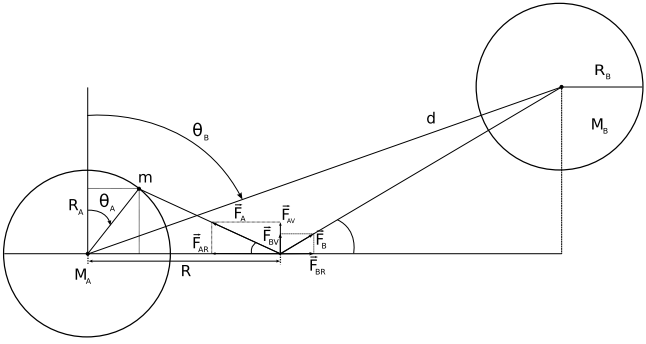}
\end{center}
\caption{\label{fig_1}Schematic picture illustrating the generation of a
  gravitational perturbing force in an equatorial accretion disc by 
  a single ``mountain'' on the surface of the neutron star 
  $(M_{\sssm{A}},R_{\sssm{A}})$ or by the binary companion
  $(M_{\sssm{B}},R_{\sssm{B}})$.}
\end{figure}

We consider here the situation illustrated in Fig.\,\ref{fig_1} with a 
basically isotropic neutron star of mass $M_{\sssm{A}}$ and radius 
$R_{\sssm{A}}$ spinning about its rotation axis with angular velocity 
$\Omega_{\sssm{A}}$ and with the symmetry plane of the accretion disc being 
orthogonal to the rotation axis. We use spherical polar coordinates 
($R,\theta,\varphi$) with the origin at the centre of the neutron star. 
Possible neutron-star asymmetry is approximated by a point-like source with 
mass $m$ located on the surface of the star at angle $\theta = 
\theta_{\sssm{A}}$. The binary companion (with mass $M_{\sssm{B}}$) is taken to 
be moving on an orbit at a constant distance $d$ from the neutron star at 
$\theta = \theta_{\sssm{B}}$. For thin discs, the angular velocity profile 
$\Omega_{\mathrm{d}}(R)$ is generally well-approximated as being 
Keplerian \citep{Nov-Tho:1973:BlaHol:} but for thick (or slim) discs, there is 
a deviation away from this because of the action of pressure 
forces \citep{Jar-Abr-Pac:1980:ACTAS:}. We here determine the radial and 
vertical components of the gravitational force produced by the perturbing 
sources in a purely Newtonian way. This is an approximation, but we do not 
expect that a relativistic analysis would greatly change the qualitative 
features of our results. For simplicity the force will be determined in the 
equatorial plane, i.e., in the symmetry plane of the disc; this is completely 
correct for thin, Keplerian discs, and gives good estimates for slim discs. We 
determine the time evolution of the perturbing force components for a fixed 
point on the disc with coordinates ($R,\theta=\pi/2, 
\varphi=\Omega_{\mathrm{d}}t$), making the restriction 
$R_{\sssm{A}}<R<10\,R_{\sssm{A}}$ since kHz QPOs are being considered. For the 
binary companion, we assume $d\gg R,R_{\sssm{B}}$.

\subsection{Neutron-star asymmetry}\label{isolmount}

Here, and in the following, we consider the force acting on a comoving unit 
mass element of the accretion disc at a given radius $R$. Using the quantities 
$x\equiv R_{\sssm{A}}/R$ and $\omega_{\sssm{A}} \equiv|\Omega_{\sssm{A}} - 
\Omega_{\mathrm{d}}|$, the vertical component of the perturbing gravitational 
force is given by
 \begin{eqnarray}
  F_{\sssm{AV}}(t) & = & \left(Gm/R^2_{\sssm{A}}\right)\,x^3 
  \cos\theta_{\sssm{A}} 
  \nonumber \\
  && \hspace{1cm}\times \left(1-2x\sin\theta_{\sssm{A}} 
  \cos\omega_{\sssm{A}}t+x^2\right)^{-3/2} ,
\end{eqnarray}
 while the radial component is given by 
\begin{eqnarray}
  F_{\sssm{AR}}(t) & = & \left(Gm/R^2_{\sssm{A}}\right)\, x^2 
  (1-x\sin\theta_{\sssm{A}} \cos\omega_{\sssm{A}} t) \nonumber \\ 
  && \hspace{1cm} \times \left(1-2x\sin\theta_{\sssm{A}} 
  \cos\omega_{\sssm{A}}t+x^2\right)^{-3/2} .
 \end{eqnarray}
 The vertical force oscillates around its mean value with amplitude
 \begin{eqnarray}
 A_{\sssm{V}} & = & \left(Gm/R^2_{\sssm{A}}\right)\,x^3 \cos\theta_{\sssm{A}} 
  \nonumber \\
  && \hspace{0.3cm} \times \left[ \left(1 - 2x\sin\theta_{\sssm{A}} + 
  x^2\right)^{-3/2} - \left(1 + x^2\right)^{-3/2} \right]
\end{eqnarray}
 and the radial force oscillates with amplitude
 \begin{eqnarray}
 A_{\sssm{R}} & = & \left(Gm/R^2_{\sssm{A}}\right)\,x^2 \left[ (1 - 
  x\sin\theta_{\sssm{A}}) \left(1 - 2x\sin\theta_{\sssm{A}} + x^2\right)^{-3/2} 
  \right. 
\nonumber \\
 && \hspace{4.3cm} \left. - \left(1 + x^2\right)^{-3/2} \right]\, .
\end{eqnarray}
 These oscillations have an anharmonic character which means that, when Fourier 
analysed, they show both the basic frequency $\omega_{\sssm{A}}$ and also some 
additional frequencies related to it.

\subsection{Binary companion}\label{binary}

The binary companion is taken to be orbiting the neutron star at a constant 
distance $d$, with angular velocity $\Omega_{\sssm{B}}$. The vertical and 
radial components of the perturbing force acting on the accreting material are 
then given by
 \begin{eqnarray}
 F_{\sssm{BV}}(t) & = & \left(GM_{\sssm{B}}/d^2\right)\cos\theta_{\sssm{B}} 
 \nonumber \\
 & & \ \times \left[1-2(R/d)\sin\theta_{\sssm{B}} 
 \cos\omega_{\sssm{B}} t + (R/d)^2\right]^{-3/2}
\end{eqnarray} 
 \begin{eqnarray}
 F_{\sssm{BR}}(t) & = & (GM_{\sssm{B}}/d^2)
 \left[\sin\theta_{\sssm{B}} \cos\omega_{\sssm{B}}t-(R/d)\right] 
 \nonumber \\
 & & \ \times \left[1-2(R/d)\sin\theta_{\sssm{B}} \cos\omega_{\sssm{B}} t + 
 (R/d)^2\right]^{-3/2},
\end{eqnarray}
 where the relative angular velocity $\omega_{\sssm{B}} = 
\left|\Omega_{\sssm{B}}-\Omega_{\mathrm{d}}\right| \approx\Omega_{\mathrm{d}}$. 
In general, these relations represent anharmonically oscillating forces but we 
are taking $R/d\ll 1$ and they then reduce to an approximate form which is 
harmonic with frequency $\omega_{\sssm{B}} \approx\Omega_{\mathrm{d}}$. The 
vertical force oscillates around its mean value with amplitude
 \begin{equation}
 B_{\sssm{V}} \approx3(GM_{\sssm{B}}/d^2)(R/d) \cos\theta_{\sssm{B}} \sin 
 \theta_{\sssm{B}}
\end{equation}
 and the radial force oscillates with amplitude
\begin{equation}
B_{\sssm{R}} \approx(GM_{\sssm{B}}/d^2)\sin\theta_{\sssm{B}}.
\end{equation}

\section{Magnitudes of the neutron-star asymmetries}\label{surfinho}

In this section we give estimates for the magnitudes of neutron-star 
asymmetries arising in different ways. We first consider classical crystalline 
mountains and magnetically-confined accretion columns; both of these are found 
to be inadequate for the present purposes, however. We then turn to some 
different observationally-motivated possibilities which seem to be more 
promising.

\subsection{Isolated crystalline mountains}\label{crystal}

Assuming that the basic nature of a mountain on the surface of a neutron star 
is the same as for mountains on planets, the pressure at the base of the 
mountain needs to be less than the maximum shear stress of the surface 
material. This pressure is given by $P_{\mathrm{mnt}} = 
\rho_{\mathrm{mnt}}g_{\mathrm{ns}}h_{\mathrm{mnt}}$, where 
$\rho_{\mathrm{mnt}}$ is the average density of the material in the mountain, 
$g_{\mathrm{ns}}$ is the surface gravity of the neutron star and 
$h_{\mathrm{mnt}}$ is the height of the mountain. The base of the mountain 
would be located at the outermost solid surface layer of the neutron star. The 
relevant density to take for this layer is rather uncertain; we will take it as 
being $\sim\! 10^6\,\mathrm{g\,cm^{-3}}$ and put $\rho_{\mathrm{mnt}}$ equal to 
that. The surface gravity is given by $g_{\mathrm{ns}} = G 
M_{\sssm{A}}/R^2_{\sssm{A}}$. For a neutron star of mass $1.4\,\msun$ and 
radius $10\,\mathrm{km}$, $g_{\mathrm{ns}}$ is $1.87 \times 
10^{14}\,\mathrm{cm\,s^{-2}}$.

Following \citet{Str-etal:1991:ASTRJ2:}, the shear modulus of the
neutron star surface material is taken to be
\begin{equation}                        
  \mu = \frac{0.1194}{1+1.781\times(100/\Gamma)^2}\frac{n(Ze)^2}{a}\,,
\end{equation}
 where $n$ is the number density of ions, $a$ is the inter-ionic distance, $Z$ 
is the atomic number of the dominant ionic species, and $\Gamma$ is the Coulomb 
coupling parameter ($\Gamma > 10^3$ for all practical purposes). The maximum 
shear strain in the surface of the neutron star has been calculated to be 
$\Theta\sim 10^{-5}\mbox{--}10^{-3}$ \citep{Smo-Wel:1970:PHYRL:}, although 
there are suggestions that it might also be as high as 
$10^{-2}$ \citep{Ush-Cut-Bil:2000:MONNR:}. Taking typical values $Z = 26$, 
$n\simeq 10^{28}\,\mathrm{cm^{-3}}$ and $a\simeq 7\times10^{-10}\,\mathrm{cm}$, 
the corresponding maximum shear stress is then $S = \mu\Theta\simeq 
10^{18}\mbox{--}10^{21}\,\mathrm{dyn\,cm^{-2}}$. The maximum height of the 
mountain is obtained by setting $P_{\mathrm{mnt}} = S$, which gives 
$h_{\mathrm{mnt}}^{\mathrm{max}}$ in the range $0.01\mbox{--}10\mathrm{cm}\ $. 
Taking the highest of these values, we then get the maximum possible mass of 
the mountain as being
 \begin{equation}
m_{\mathrm{mnt}}\sim\rho_{\mathrm{mnt}} 
(h^{\mathrm{max}}_{\mathrm{mnt}})^3 \sim 10^9\,\mathrm{g}\sim 
10^{-24}\mbox{--}10^{-25}\,M_{\sssm{A}}
 \end{equation}
 which is too small to be relevant here.

\subsection{Accretion columns}\label{accrecol}

For a neutron star with a strong magnetic field, accreting matter close to it 
can be diverted away from the equatorial plane and form accretion columns above 
the magnetic poles \citep{Wos-Wal:1982:ApJ:,Ham-etal:1983:AA:}. If the amount 
of matter in the columns is sufficiently large, this can provide another source 
of neutron-star asymmetry which could again be modelled in terms of 
``mountains'' (probably two symmetric ones in this case). This has been invoked 
in connection with gravitational-wave emission \citep{Mel-Pay:2005:ASTRJ2:} and 
we check here whether it could also be relevant in the present context.

The column height can be determined from the condition that the flow will start 
to spread out sideways when the pressure of the matter in the column 
$P_{\mathrm{ac}}$ becomes large enough to bend the magnetic field lines 
outwards, typically when it is about a hundred times greater than the confining 
magnetic pressure, i.e. when 
 \begin{equation}
  P_{\mathrm{ac}} \sim 4 \times 10^{24}\,\mathrm{dyn\,cm}^{-2}\,
  \left(B_{\mathrm{s}}/10^{12}\,\mathrm{G}\right)^2, 
\end{equation}
 where $B_{\mathrm{s}}$ is the strength of the surface dipole field 
\citep{Bro-Bil:1998:ASTRJ2:}. For having hydrostatic equilibrium, this pressure 
should be the same as that elsewhere at the same level in the crust. In the 
density range $10^6\,\mathrm{g\,cm^{-3}} \leq \rho \leq 
10^{10}\,\mathrm{g\,cm^{-3}}$ the pressure can be expressed using the fitting 
formula $\log P = 13.65 + 1.45 \log \rho$ \citep{Bay-Pet-Sut:1971:ASTRJ2:} and 
the relation between the field strength and the density at the bottom of the 
column is then given by $(\rho_{\mathrm{bot}}/10^6\,\mathrm{g\,cm^{-3}}) \sim 
36 (B_{\mathrm{s}}/10^{12}\,\mathrm{G})^{1.38}$. Using this, the scale height 
of the column, $h_{\mathrm{ac}}$ is then
 \begin{equation}
  h_{\mathrm{ac}} \sim P_{\mathrm{ac}} / (\rho_{\mathrm{bot}}\,
    g_{\mathrm{ns}})
    \sim 10^3 \mathrm{cm}
    \left( {B_{\mathrm{s}}}/{10^{12}\,\mathrm{G}}\right)^{0.62}\, .
\end{equation}
 Following \citet{Sha-Teu:1983:BHWDNS:}, the cross-sectional area of the column 
is estimated as
\begin{eqnarray}
  A_{\mathrm{xs}} & \sim & 10^{10}\,\mathrm{cm^2}
    \left(\frac{B_{\mathrm{s}}}{10^{12}\,\mathrm{G}}\right)^{-4/7}
    \left(\frac{M_{\sssm{A}}}{1.4\,\msun}\right)^{1/7} \nonumber \\
    && \hspace{1.5cm} \times 
    \left(\frac{R_{\sssm{A}}}{10^6\,{\mathrm{cm}}}\right)^{9/7}
    \left(\frac{\mdot}{10^{-9}\,\msun/{\mathrm{yr}}}\right)^{2/7}\,,
\end{eqnarray}
 where $\mdot$ is the accretion rate, which we normalise to a typical value for 
LMXBs. The mass of the column is then given by
 \begin{eqnarray}
  m_{\mathrm{ac}} & \sim & \rho_\mathrm{bot} A_{\mathrm{xs}} h_{\mathrm{ac}}
    \nonumber \\
  & \sim & 10^{-13}\,M_{\sssm{A}}
       \left(\frac{B_\mathrm{s}}{10^{12}\,\mathrm{G}}\right)^{10/7}
       \left(\frac{M_{\sssm{A}}}{1.4\,\msun}\right)^{-6/7} \nonumber \\
       && \hspace{1.5cm} \times 
       \left(\frac{R_{\sssm{A}}}{10^6\,\mathrm{cm}}\right)^{9/7}
       \left(\frac{\mdot}{10^{-9}\,\msun/\mathrm{yr}}\right)^{2/7}\,
\end{eqnarray}
 which is again very small, even for a field as high as $10^{12}\,\mathrm{G}$. 
The continued existence of a roughly Keplerian accretion disc down to small 
radii, as required for our QPO picture, needs the magnetic field to be 
suitably low and so magnetically-confined accretion columns do not seem to 
be relevant for the present purposes.

\subsection{Quadrupole moments inferred from limiting neutron star 
spin rates}\label{quadrupole}

\citet{Bil:1998:ASTRJ2:} noted that the spin frequencies for many 
accreting weakly-magnetised neutron stars were thought to lie in a 
rather narrow range around $300\,\mathrm{Hz}$, which is a much lower 
frequency than that corresponding to centrifugal break-up. Since these 
objects are thought to have been accreting for long enough so as to 
gain sufficient angular momentum to reach the break-up limit, it 
seemed that some mechanism was halting the spin-up. Despite the fact 
that the spread of spin frequencies is now thought to be less peaked 
\citep[see][]{Man:2005:AJ:}, the issue of why spin-up should be halted 
before the break-up limit is reached remains a relevant one. Bildsten 
suggested that this might be caused by the accretion torque becoming 
balanced by a gravitational-wave torque resulting from an asymmetry of 
the neutron star. Taking the asymmetry to be represented by an $l=m=2$ 
perturbation, one finds that the magnitude of the misaligned 
quadrupole moment required for attaining this equilibrium at a 
frequency $\nu_{\mathrm{s}}$ is given by
 \begin{eqnarray}
  Q_{\mathrm{eq}} & = & 3.5 \times 10^{37}\,{\mathrm{g\,cm}}^2 
    \left(\frac{M_{\sssm{A}}}{1.4\,\msun}\right)^{1/4}
    \left(\frac{R_{\sssm{A}}}{10^6\,{\mathrm{cm}}}\right)^{1/4} \nonumber \\
    & & \hspace{1.2cm} \times 
    \left(\frac{\mdot}{10^{-9}\,\msun/{\mathrm{yr}}}\right)^{1/2}
    \left(\frac{\nu_{\mathrm{s}}}{300\,\mathrm{Hz}}\right)^{-5/2}\, . 
\end{eqnarray}
 In terms of our simplified model of representing the asymmetry by means of 
point masses on the surface of an otherwise spherical neutron star, this 
corresponds to
 \begin{eqnarray}
  m_{\mathrm{quad}} & \sim & 10^{-8}\,M_{\sssm{A}}\, 
    \left(\frac{M_{\sssm{A}}}{1.4\,\msun}\right)^{-3/4}
    \left(\frac{R_{\sssm{A}}}{10^6\,{\mathrm{cm}}}\right)^{-7/4} \nonumber \\
    & & \hspace{0.9cm} \times 
    \left(\frac{\mdot}{10^{-9}\,\msun/{\mathrm{yr}}}\right)^{1/2}  
    \left(\frac{\nu_{\mathrm{s}}}{300\,\mathrm{Hz}}\right)^{-5/2}\, .
\label{mquad}
\end{eqnarray}
 Various mechanisms have been suggested in the literature for 
producing such a value (or even higher), involving global deformations 
of the neutron star rather than an isolated mountain on the surface 
\citep[][, etc.]{Bil:1998:ASTRJ2:, Ush-Cut-Bil:2000:MONNR:, 
Has-Jon-And:2006:MONNR:}. We will not enter into details of this here 
but we note that values of the deformation large enough to be capable 
of explaining the limitation of neutron star spin-up, as suggested by 
Bildsten, may also be large enough for the present purposes: in the 
next section, we take the value for the point mass given by 
Eq.\,(\ref{mquad}) with the canonical parameter values, and check 
whether this same value $\sim\!10^{-8}\,M_{\sssm{A}}$ might 
\emph{also} be large enough to be relevant for inducing the QPO 
behaviour.

\section{Discussion and conclusions}\label{compar}

Here, we discuss whether the effect of the neutron-star asymmetries or 
of the binary companion could be large enough to account for 
excitation of the QPO phenomenon. In our simplified picture (c.f. 
Eq.\,(\ref{resonance})), the excitation time for the amplitude $a$ of 
resonant oscillations to grow to a particular value is given by
 \begin{equation}
  t_{\mathrm{ex}} = \left(\frac{a}{R}\right)\,
    \left(\frac{\alpha}{\pi}\right)\,
    \left(\frac{f_{\mathrm{p}}}{f_0}\right)^{\!-1}\,
    \tau_{\sssm{K}}\,,
\end{equation}
 where $f_p$ is the amplitude of the perturbing force acting on unit 
mass, $f_0 = GM_{\sssm{A}}/R^{2}$ is the main gravitational force from 
the central object, $\tau_{\sssm{K}} = 2\pi/\Omega_{\sssm{K}}$ is the 
period of circular Keplerian motion at the location being considered 
and the epicyclic frequency of the perturbation being excited is 
$\omega = \alpha\Omega_{\sssm{K}}$. The dimensionless radial and 
vertical ``epicyclic functions'' satisfy $\alpha \le 1$ everywhere. 
Note that since $\tau_{\sssm{K}} \sim 10^{-3}\,\mathrm{s}$, the 
amplitude amplification in 1 second is $\sim\! 10^{3}$. The ratio 
$a/R$ needs to grow to $> 10^{-3}$ in order to potentially explain the 
QPO behaviour and it would need to do that within 
$\sim\!10^{3}\,\mathrm{s}$ to account for the QPO phenomena seen in 
atoll sources.

In the case of a binary companion, for the radial perturbing force one has
 \begin{eqnarray}
  \frac{f_{\mathrm{p}}}{f_0} & \sim & 
    \left(\frac{M_{\sssm{B}}}{M_{\sssm{A}}}\right)
    \left(\frac{R}{d}\right)^{2} \nonumber \\
    & \sim & 10^{-9} \left( \frac{M_{\sssm{B}}}{0.1\,M_{\sssm{A}}} \right)
      \left( \frac{R}{10^{6}\,\mathrm{cm}}\frac{10^{10}\,\mathrm{cm}}{d}
      \right)^{2}\, ,   
\end{eqnarray}
 where we have normalised to typical parameter values. This could 
produce $a/R \sim 10^{-3}$ at times $t_{\mathrm{ex}}\lesssim 
10^{3}\,\mathrm{s}$, but there is a problem for it producing 
resonances in the innermost parts of the disc because of having 
$\omega_{\sssm{B}} \approx\Omega_{\sssm{K}}$. The corresponding 
vertical force is smaller by at least four orders of magnitude and is 
clearly irrelevant. (Note that the vertical force is a tidal force 
whereas the radial one is a direct gravitational attraction; also, 
$\theta_{\sssm{B}}$ is probably rather close to $\pi/2$.)

For the neutron star asymmetries, we focus on the case of the 
misaligned quadrupole moments where $f_{\mathrm{p}}/f_0$ can be $\sim 
10^{-8}$ in the inner parts of the disc for both radial and vertical 
oscillations (taking $m_{\mathrm{quad}} = 10^{-8}\,M_{\sssm{A}}$). For 
this, $a/R$ could reach $10^{-3}$ in $\lesssim 10^2 \,\mathrm{s}$, 
which encourages further investigation of this scenario.

We conclude that at least one of the types of gravitational 
perturbation considered in this paper might provide a plausible 
mechanism for inducing kHz QPO behaviour although many details remain 
to be worked out (in particular concerning the response of the fluid 
medium and the production of the luminosity variations). The influence 
of the binary companion could possibly be effective in providing the 
perturbations but the more likely possibility is that they might be 
produced by the neutron-star asymmetries. It is striking that the same 
magnitude for the misaligned quadrupole moment as advocated elsewhere 
for explaining limiting neutron star periods, seems also to give a 
plausible mechanism for inducing QPO behaviour.

\vspace{\baselineskip}

\hrule

\vspace{\baselineskip}

\section*{Acknowledgements}

This work was supported by Czech grants MSM~4781305903 and
GA\v{C}R~202/06/0041.

\bibliography{7061}

\end{document}